# Atomic and electronic structure of poly-[Ni(Salen)]: combined study by XPS, UV PES, NEXAFS and DFT methods

Petr M. Korusenko[1,2,3], Olga V. Petrova [1,4], Anatoliy A. Vereshchagin[2], Oleg V. Levin[2], Ratibor G. Chumakov[5], Konstantin P. Katin[6], Sergey V. Nekipelov[4], Victor N. Sivkov[4], Alexandra V. Koroleva[7], Alexander S. Konev[2] and Alexander S. Vinogradov[1]

[1] V.A. Fock Institute of Physics, St. Petersburg State University, 7/9 Universitetskaya nab., 199034 Saint Petersburg, Russia

[2] Institute of Chemistry, St. Petersburg State University, 7/9 Universitetskaya nab., 199034 Saint Petersburg, Russia

[3] Omsk State Technical University, 11 Mira prosp., 644050 Omsk, Russia

[4] Institute of Physics and Mathematics, Komi Science Centre, Ural Branch of the Russian Academy of Sciences, 167982 Syktyvkar, Russia

[5] Kurchatov Synchrotron Radiation Source, National Research Center Kurchatov Institute, Moscow 123182, Russia

[6] National Research Nuclear University "MEPhI", Kashirskoe Sh. 31, 115409 Moscow, Russia

[7] Research Park, St. Petersburg State University, 7/9 Universitetskaya nab., 199034 Saint Petersburg, Russia

**Abstract**

A detailed study of poly-[Ni(Salen)] polymer in its oxidized (Ox) and reduced (Red) states was conducted using X-ray photoelectron (XPS) and ultraviolet photoemission (UV PES) spectroscopy, near-edge X-ray absorption fine structure (NEXAFS) spectroscopy, and quantum-chemical calculations. XPS analysis revealed significant energy shifts (-1.5 to -1.8 eV) and broadening of the PE lines for all atoms upon polymerization, indicating a major redistribution of valence electron density between the monomer fragments. In the oxidized polymer, new features in the Ni $2p$ and O $1s$ PE spectra were associated with the formation of polarons with weakened Ni–O bonds; this effect diminished upon reduction as the number of polarons decreased. Quantum-chemical calculations attributed the valence band broadening to enhanced C $2p$ contributions from π-conjugation between monomers. NEXAFS spectroscopy confirmed the stability of the ethylenediamine fragment and the direct involvement of the phenolic rings of the salen ligand in polymerization, also revealing a partial weakening and incomplete restoration of the π bonding between O and Ni atoms upon reduction. Furthermore, it was shown that it is the $BF_4^-$ anions that weaken the Ni–O bonds during oxidation, which are partially preserved in the reduced state.

# 1. Introduction

Functional materials obtained by polymerization of transition metal complexes with salen-type ligands [M(Salen)] (M = Ni, Cu, Co, Fe, V, Pd, etc.) have high redox conductivity, exhibit luminescent ability and magnetic properties and are considered as promising materials for a wide range of applications from catalytic systems to chemical power sources [1–12].

It is known that the process of electrochemical polymerization of the [M(Salen)] complexes is carried out in two stages [13,14]. At the first stage, the molecular complexes are sorbed from the electrolyte solution with the formation of a surface layer on the working electrode. At the second stage, after applying the oxidation potential to the electrode with adsorbed layers, reactions occur between the [M(Salen)] monomers, followed by their covalent cross-linking and the formation of a polymer film. To understand the mechanisms of formation of poly-[M(Salen)] polymers and to describe their electronic properties, detailed knowledge of the atomic-electronic structure in both the charged (oxidized) and neutral (reduced) states of these compounds is necessary. However, these data are currently limited, making it difficult to fully understand and accurately describe the electronic properties of poly-[M(Salen)] polymers. At the same time, this important information can be obtained using the methods of modern X-ray spectroscopy: X-ray photoelectron (XPS) and valence band photoemission (VB PE) spectroscopy, as well as Near-Edge X-ray Absorption Fine Structure (NEXAFS) spectroscopy [15,16].

It is worth noting that XPS is the most widely used method for characterizing energy levels of electrons in the filled inner (core) electron shells of individual atoms in polyatomic systems [15]. Core-level photoemission (CL PE) spectra are excited by monochromatic X-ray lines and are employed to measure the binding energy ($E_{bin}$) of the core electrons and their changes (chemical shifts) in various compounds. In turn, VB PES is a key method for studying the electronic structure of the valence band. It uses ultraviolet (UV) and soft X-ray photons to excite VB PE spectra, which reflect the energy distribution of occupied electronic states in the valence band and provide information about the atomic orbital (AO) composition of the molecular orbitals (MOs) responsible for these valence states.

Finally, NEXAFS spectra provide valuable information about the low-energy unoccupied electronic states of the polyatomic system under study. The structure of these spectra in the vicinity of the ionization threshold of the inner (core) electron shell of the absorbing atom is characterized by absorption bands (resonances) that appear as a result of dipole-allowed transitions of core electrons to final unoccupied electron states. The intra-atomic nature of these absorption processes ensures probing of these states in the region of the absorbing atom and the atoms of its nearest surroundings. The spatially localized nature of the X-ray excitations of the atom in a polyatomic system allows the final electronic states to be approximated using the one-electron MOs of a fragment or otherwise quasi-molecule, formed by the absorbing atom and its nearest surroundings atoms [17].

It is important to note that studies in recent years are characterized by some attempts to use XPS, VB PES, NEXAFS spectroscopy to characterize the atomic and electronic structure of [M(Salen)] monomers and their polymers [8–10,12,18–35]. Moreover, almost all the measurements of the XPS spectra were performed on laboratory spectrometers with low energy resolution and insufficiently good statistics [9,19,21–24,28,31]. At the same time, the NEXAFS spectra of the functional nitrogen, oxygen, and carbon atoms in poly-[M(Salen)] polymers, as well as VB PE spectra of occupied electron states of the valence band are still unknown.

In previous work [35], the atomic-electronic structure of the $[NiO_2N_2]$ coordination center in the [Ni(Salen)] complex and its polymer poly-[Ni(Salen)] was studied using NEXAFS and EXAFS spectroscopy in the vicinity of Ni 1$s$ and 2$p$ absorption edges along with quantum chemical calculations for these polyatomic systems. It was demonstrated that during electrochemical oxidation of the monomers, the coordination center undergoes a structural distortion, and, then, upon transition to the reduced (neutral) state, its structure returns almost completely to the square-planar geometry. In addition to these geometric changes, it was also established that the basic structural unit for the complex in the condensed state is a d-d-stacking dimer. In turn, a tetramer formed by cross-linking a pair of such dimers is the structural unit for the polymer. However, several important aspects of the polymerization process still remain not fully understood. These include: (i) the electron density redistribution (direction and magnitude of the electron transfer) between ligand atoms and the coordinating Ni cation during the polymerization process, (ii) the chemical state of the counterions adsorbed from the electrolyte on the poly-[Ni(Salen)] surface, and (iii) the effect of these counterions on the electronic properties of the polymer.

In view of the above, this study aims to investigate the electronic structure of the polymer poly-[Ni(Salen)] by both the experimental and theoretical methods in order to determine: (i) the charge (chemical) state of the coordinating nickel atom, the ligand atoms (O, N and C) and the atoms of the counterions adsorbed by the polymer; (ii) the atomic-orbital composition of occupied and unoccupied electronic states. For this purpose, a comparative analysis of the measured XPS, UV PE and NEXAFS spectra was performed for the molecular complex [Ni(Salen)] and its polymer in different (oxidized and reduced) states. The identification of the AO contributions of the ligand and nickel atoms to the occupied MOs responsible for the main bands in the UV PE spectra of the valence band of poly-[Ni(Salen)] was performed based on an empirical analysis of the obtained spectra of the complex and polymer, and supplemented by DFT calculations of the electronic structure of the valence band of structural fragments characterizing these systems.

## 2. Materials and Methods

*2.1 Sample preparation*

The powder of molecular complex [Ni(Salen)] was synthesized using a conventional approach, involving the reaction of nickel acetate (Ni(OAc)$_2$) with H$_2$(Salen) in an ethanol solution [36, 37]. The crude product was purified via recrystallization in ethanol and dried under vacuum at 80 °C. Poly-[Ni(Salen)] films were electrodeposited onto a platinum (Pt) electrode by cyclic voltammetry (CV) using an Autolab PGSTAT30 potentiostat. The electropolymerization process employed a 0.001 M monomer solution dissolved in a 0.1 M LiBF$_4$/CH$_3$CN electrolyte. To investigate the effect of the adsorbed counterions on the charge state of the coordination center atoms of the monomeric fragments in the polymers, additional polymeric films were synthesized using a 0.1 M LiClO$_4$/CH$_3$CN solution as the electrolyte. A three-electrode cell was utilized, comprising (i) a Pt working electrode, (ii) an MF-2062 reference electrode (Bioanalytical Systems) filled with 0.1 M Et$_4$NBF$_4$/CH$_3$CN containing 0.001 mol/L AgNO$_3$, (iii) a stainless-steel mesh counter electrode. To produce polymers in different charge states, polymerization was halted at predefined potentials: −0.2 V for the reduced (Red) state and +1.0 V for the fully oxidized (Ox) state. The ~420 nm thick films were obtained after 15 polymerization cycles at a scan rate of 50 mV/s. All synthesis processes were conducted in an argon-atmosphere glovebox, and final products were stored in Eppendorf tubes to minimize interaction with the external environment and maintain their stability.

*2.2 Methods*

*2.2.1 XPS and UV PES*

The [Ni(Salen)] samples for study were prepared as thin polycrystalline layers (up to 30 nm) on a Ti substrate *in situ* by thermal evaporation of dehydrated powder from a quartz crucible. Details of the preparation of [Ni(Salen)] are presented in [26]. In the case of poly-[Ni(Salen)], as-prepared polymer layers were used as measurements samples after being preheated *in situ* for 30 minutes at a temperature of about 190 °C. A LiBF$_4$ powder rubbed into a scratched Pt substrate was used as a reference. The overview and CL (Ni 2$p$, O 1$s$, C 1$s$, N 1$s$) PE spectra of [Ni(Salen)] and its poly-[Ni(Salen)] polymer in different charge (reduced, oxidized) state as well as F 1$s$ and B 1$s$ PE spectra of counterions BF$_4^-$ adsorbed by the polymer were measured at the Research park of St. Petersburg State University (Centre for Physical Methods of Surface Investigation, St. Petersburg, Russia), using an ESCALAB 250Xi laboratory electron spectrometer (Thermo Fisher Scientific, Waltham, MA, USA). All the PE spectra were registered in the normal photoemission geometry and the angle-integrated mode, relative to the Fermi level of the sample. These spectra were recorded using monochromatized AlKα radiation ($h\upsilon$ = 1486.6 eV) in the constant analyzer energy (CAE) mode at pass energies of 50 eV and 10 eV, respectively. In the case of F 1*s* and B 1*s* spectra, their registration was carried out at pass energy 50 eV. The total energy resolution of the measured CL PE spectra was 800 meV. The $E_{bin}$ scale was calibrated against the Pt 4$f_{7/2}$ PE peak ($E_{bin}$ = 71.1 eV) measured from a clean platinum foil. The charging of the samples under X-ray irradiation was eliminated by a compensation system, using irradiation of the samples with a low-energy electron flux. The experimental data were processed using the Thermo Avantage ver. 5.9931 software package. A detailed CL PE peak analysis was performed by peak fitting using the Gaussian/Lorentzian product formula from the CasaXPS 2.3.16 software [38].

The photons of the He(I) resonance line (*hv* = 21.2 eV) were used to excite the UV PE spectra of the [Ni(Salen)] and its polymer. These PE spectra were obtained at an analyzer pass energy of 1.0 eV, which provided a total energy resolution of 360 meV. When recording the UV PE spectra, the system of electron–ion charge compensation was not used due to technical limitations. All PE measurements were carried out at room temperature and base pressure around 2·10$^{-10}$ mbar.

*2.2.2 NEXAFS*

NEXAFS experiments were carried out using monochromatic synchrotron radiation (SR) and facilities of the Russian–German beamline (RGL-PES end-station) at the BESSY II electron storage ring (Berlin, Germany) [39] as well as a "NanoPES" end-station at the "KISI-Kurchatov" SR source [40]. The [Ni(Salen)] sample and the poly-[Ni(Salen)] polymer films for the study were prepared in the same way as in the case of XPS measurements (see 2.2.1 section). LiBF$_4$ powder mechanically rubbed into a scratched indium substrate was as a reference sample. The NEXAFS spectra were measured under identical conditions using both end-stations and recorded in the total electron yield (TEY) mode by detecting sample drain current, with the photon incident angle of 45°. Photon energy resolution in the vicinity of Ni 2$p_{3/2}$ (~ 850 eV), F 1*s* (~ 685 eV), O 1*s* (~ 530 eV), N 1*s* (~ 400 eV), C 1*s* (~ 280 eV) and B 1*s* (~ 190 eV) absorption edges was 300, 280, 150, 90, 60 and 50 meV, respectively. All NEXAFS spectra were normalized to the incident photon flux, which was monitored by recording photocurrent from a clean gold mesh. Photon energy was calibrated by measuring Pt 4$f_{7/2}$ PE signals excited by radiation in the first and second diffraction orders and taking their difference. The measurements were carried out at room temperature and base pressure around 2·10$^{-10}$ mbar. No noticeable effects of charging or deterioration of the [Ni(Salen)] and poly-[Ni(Salen)] samples irradiated by intense synchrotron beam were observed.

*2.3 DFT calculation details*

The atomic and electronic structure calculations for the [Ni(Salen)] complex and its polymers were performed by the DFT method using the GAMESS-US package [41]. The d-d-dimer, previously defined in [35], as well as the tetramer, consisting of these cross-linked dimer blocks, were used as structural fragments modeling the complex in the condensed phase and its polymers, respectively. The geometry of these structures was optimized using the B3LYP functional [42,43]. For the d-d dimer, the total charge (q) during optimization was 0, while for the tetramer, optimization was performed at q = 0 and q = +3 (with one $BF_4^-$ counterion) to simulate the reduced and oxidized states of the polymer, respectively (Figure 1). Electronic multibasis was constructed from 311+G** basis functions of organic ligands and lanl2tz+ basic functions of Ni atoms [44]. The geometry of all structures was optimized without any restrictions on symmetry, and the residual forces did not exceed $10^{-4}$ eV/A. For the analysis of the predicted density of states (DOS), all carbon atoms were divided into five groups as was done in [25]. For all the plots of the DOSs of the valence MOs, the energy levels were broadened by Gaussian lines with a full width at a half-maximum (FWHM) of 1 eV.

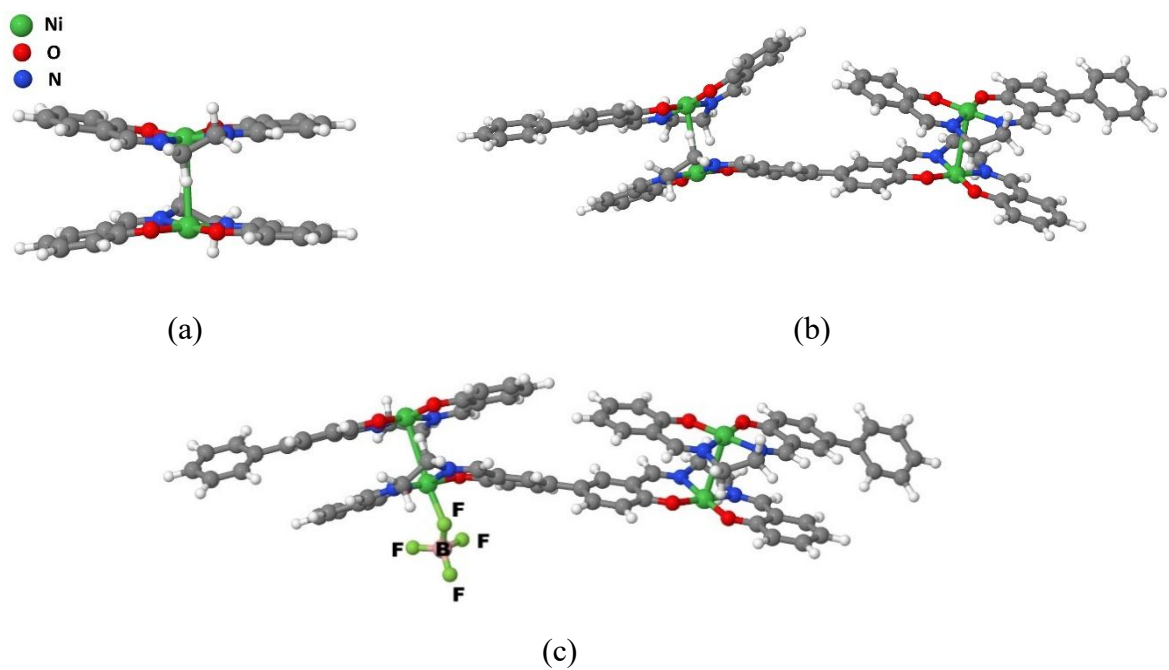

**Figure 1.** DFT optimized structures of the d-d dimer with total charge q=0 (a) and tetramers: with q=0 (b), and q=+3 (with the $BF_4^-$ counterion) (c). Carbon and hydrogen atoms are represented by gray and white balls, respectively

**Results and Discussions**

**3.1 XPS**

Let us first consider the overview PE spectra of the [Ni(Salen)] complex and its polymer in the reduced and oxidized states to characterize the quality of the prepared samples (Figure 2). It is clearly seen that the spectrum of the complex contains only the PE and Auger lines of nickel (Ni 2$s$, Ni 2$p$, Ni *LMM*, Ni 3$s$, Ni 3$p$), oxygen (O 1$s$, O *KLL*), nitrogen (N 1$s$, N *KLL*) and carbon (C 1$s$, C *KLL*). In the case of polymer films, the PE lines of boron (B 1$s$) and fluorine (F 1$s$) is additionally observed. The presence of the last two elements is due to the existence of $BF_4^-$ counterions on the surface of the polymer films, which are adsorbed from the electrolyte to compensate for the positive charge arising during the electrochemical polymerization of [Ni(Salen)] molecules [10]. It is evident from the data in Table 1 that the atomic concentration of F and B in the polymer in the oxidized state is significantly higher than in the reduced state, which is consistent with doping of the polymer film by $BF_4^-$ during charging (oxidation) and, accordingly, de-doping upon transition to the neutral (reduced) state. The presence of tetrafluoroborate fragment in the reduced state is most probably due to the absorbed electrolyte salt, which also contains tetraethylammonium cations increasing the atomic percentage of C and N in the reduced polymer as compared to the oxidized polymer. The notable deviation from the stochiometric ratio of 1:4 in $BF_4$ is due to the low signal-to-noise ratio for boron signal. For other elements, the errors in atomic percentage are within 4% typical of XPS method. In general, the presented data on the chemical composition of the samples indicate the absence of foreign inclusions in the surface layers of [Ni(Salen)] and the prepared polymer films.

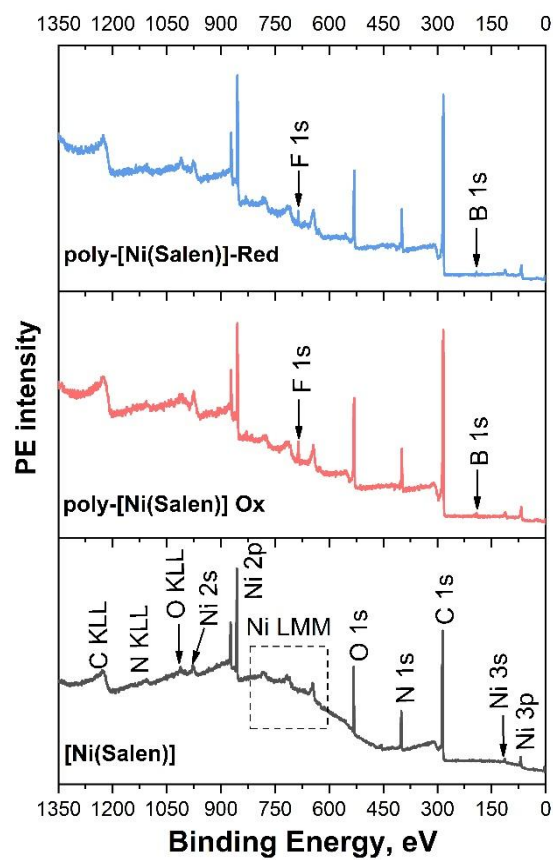

**Figure 2.** Overview PE spectra of the [Ni(Salen)] complex and its polymer poly-[Ni(Salen)] in the oxidized (Ox) and reduced (Red) states ($hv$ = 1486.6 eV)

**Table 1.** Chemical composition of *in situ* prepared [Ni(Salen)] layers onto Ti and *ex situ* prepared polymer films onto Pt according to XPS data

| Sample | Concentration, at. % | | | | | |
|---|---|---|---|---|---|---|
| | [C] | [N] | [O] | [Ni] | [B] | [F] |
| [Ni(Salen)] | 72.3 | 11.0 | 10.5 | 6.2 | - | - |
| poly-[Ni(Salen)]-Ox | 73.2 | 8.3 | 12.7 | 2.8 | 1.2 | 1.8 |
| poly-[Ni(Salen)]-Red | 75.7 | 9.1 | 10.2 | 2.9 | 0.8 | 1.3 |

Let us consider the CL (C 1*s*, N 1*s*, O 1*s*, Ni 2*p*) PE spectra of the [Ni(Salen)] complex and its polymer poly-[Ni(Salen)] in different charge (oxidized and reduced) states to determine the chemical state of the ligand atoms (carbon, nitrogen and oxygen) and nickel atoms in these systems, and let's start with a detailed examination of the C 1*s* PE spectra (Figure 3).

As can be seen from Figure 3a, the C 1*s* PE spectrum of the [Ni(Salen)] complex consists of two bands (low-energy Cα and high-energy Cβ) with maxima at binding energies of 285.8 and 287.3 eV, respectively. It has been previously shown [25,26] that the Cα band corresponds to carbon atoms of the phenyl rings of salen that are bonded only to other carbon atoms (in Figure 3b, these atoms are designated as C1 and C4). While the Cβ band corresponds to carbon atoms bonded, in addition to carbon, to oxygen and nitrogen atoms in the phenolic and ethylenediamine groups, respectively. In Figure 3b, these atoms are designated as C2, C3, and C5.

The spectra of the polymer in both the oxidized and reduced states also show the well-resolved low- and high-energy Cα and Cβ bands. Moreover, the energy distance Δ between these bands is the same for both polymer and complex and amounts to 1.5 eV. The main differences between the spectra are the -1.8 eV shift of the Cα and Cβ lines, as well as the changes in their relative intensities and widths upon going from [Ni(Salen)] to poly-[Ni(Salen)] (see Tables 2 and 3). Obviously, these findings reflect the redistribution of the valence electron density of C atoms in the monomer due to changes in their chemical bonding during polymerization. Thus, the observed low-energy shift of the C 1$s$ spectrum can be attributed to an increase in the electron density on the carbon atoms of monomeric fragments. A similar low-energy shift was observed upon going from a monomeric fragment of methacrylate to its oligomer consisting of 41 monomers [45].

The approximation parameters of the compared spectra are presented in detail in Table 2. From these data it follows that when moving from the complex to the polymer in the oxidized state, the ratio of the peak intensities of the Cα and Cβ bands changes slightly from 1.0:0.63 to 1.0:0.61. However, the ratio of the total intensities of these bands for the oxidized poly-[Ni(Salen)], on the contrary, is 1.0:0.68. This change is due to an increase in the FWHM value of the Cβ component (from ~1.0 to 1.50 eV). It is likely that individual Cβ (C2, C3, C5) atoms change their valence electron density to varying degrees during polymerization, which causes the low-energy shift of the PE line and its broadening. At the same time, upon transition from the oxidized to the reduced state, the ratio of the total intensity of the Cα to Cβ component decreases from 1.0:0.68 to 1.0:0.5, respectively. To explain this observation, consider Figure 3b. It can be seen that 8 out of the 16 carbon atoms of the monomeric fragment of the complex are of the C1 type, which contribute predominantly to the Cα band. Therefore, the notably low Cα/Cβ ratio in the spectrum of reduced poly-[Ni(Salen)] is likely attributable to an increase in the number of C–C bonds. This increase arises from the cross-linking of individual dimeric fragments of the complex via the C1 atoms of the phenyl rings during the polymerization process. In addition, when moving from the oxidized to the neutral state, a decrease in the FWHM of the Cβ component is observed from 1.50 to 1.35 eV, which can also affect the Cα/Cβ ratio.

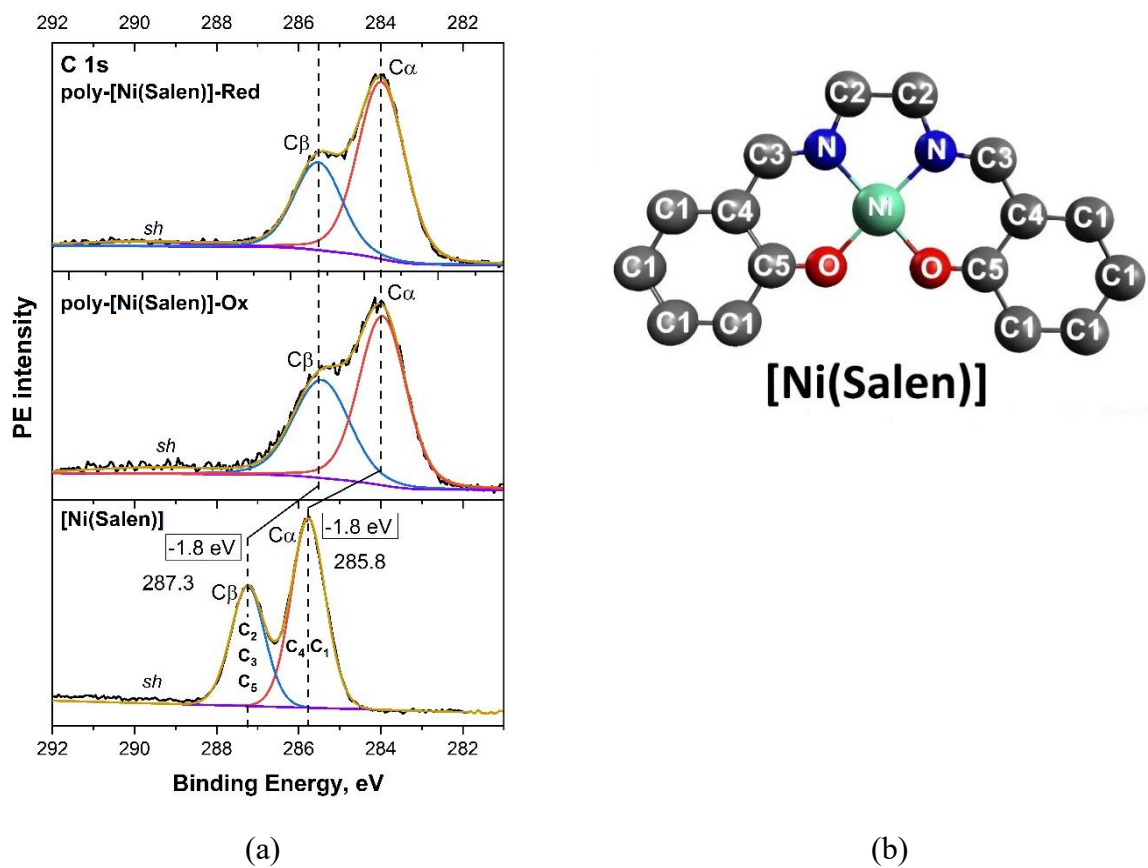

(a)                               (b)

**Figure 3.** (a) C 1$s$ PE spectra of [Ni(Salen)] and its polymer in different charge (oxidized (Ox) and reduced (Red)) states ($hv$ = 1486.6 eV). (b) Schematic view of the monomeric complex of [Ni(Salen)], in which all carbon atoms are divided into two groups C1, C4 (Cα) and C2, C3, C5 (Cβ) with atoms in similar chemical states, as was done in [25]

**Table 2.** Fitting parameters for experimental C 1$s$ PE spectra of [Ni(Salen)] and its polymer in different charge (oxidized (Ox) and reduced (Red)) states

| Sample | PE band | $E_{bin}$, eV | FWHM, eV | Relative peak intensity | Relative total intensity |
|---|---|---|---|---|---|
| [Ni(Salen)] | Cα | 285.8 | 0.97 | 1.0 | 1.0 |
|  | Cβ | 287.3 | 0.97 | 0.63 | 0.63 |
| poly-[Ni(Salen)]-Ox | Cα | 284.0 | 1.35 | 1.0 | 1.0 |
|  | Cβ | 285.5 | 1.50 | 0.61 | 0.68 |
| poly-[Ni(Salen)]-Red | Cα | 284.0 | 1.35 | 1.0 | 1.0 |
|  | Cβ | 285.5 | 1.35 | 0.49 | 0.50 |

Let us move on to examining the Ni 2$p$ PE spectra of the complex and polymer in Red and Ox states, presented in Figure 4. These spectra consist of two spin-doublet components – 2$p_{3/2}$ and $2p_{1/2}$, and have high-energy satellite bands (*sh*) located at a distance of about 4–6 eV from the main lines. It is clearly seen that the primary differences between the Ni $2p_{3/2}$ spectra of the studied systems are represented by a -1.5 eV shift of the PE lines and their broadening during the transition from the complex to the polymer. Interestingly, in the Ni $2p_{3/2}$ spectra of the polymer, a clear shoulder is observed on the low-energy side, the intensity of which is noticeably higher in the oxidized polymer compared to the reduced one. This observation suggests that the nickel atom in the polymer layer exists in two different chemical states (Figure 4, Table 3). The most intense component Ni1 is associated with nickel atoms in the [$NiO_2N_2$] coordination centers. The Ni2 component at 853 eV, on the other hand, appears to correspond to nickel atoms that have a reduced effective positive charge due to electron transfer from $BF_4^-$ counterions, which is consistent with our previous results in [35].

    The remaining differences in the Ni $2p_{3/2}$ spectra of the studied systems are associated with changes in the intensity of the *sh* bands. These bands probably arise during Ni $2p$ photoionization process as a result of the simultaneous shake-up of the electron subsystem of the atom, accompanied by charge transfer from the metal atom to the atoms of the salen ligand, which was previously proposed for the similar molecular complex [Cu(Salen)] [46]. It is clearly seen that on going from the Ni $2p$ spectrum of the [Ni(Salen)] complex to that of the polymer in the oxidized state, an increase in the intensity of these bands is observed. Earlier [47,48], a relationship was found between the intensity of *sh* satellites in Ni $2p$ photoelectron spectra of Ni(II) complexes with Schiff bases and the magnitude of the magnetic moment of the latter. It was found that paramagnetic nickel complexes exhibit a significant increase in the intensity of the *sh* satellite relative to the main Ni $2p_{3/2}$ line compared to their diamagnetic counterparts, which correlates with their higher magnetic susceptibility. In view of this, it is logical to assume that the presence of more intense satellite bands in the Ni $2p$ spectrum of poly-[Ni(Salen)]-Ox is caused by the formation, during the electrochemical oxidation of the complex, of charged particles (polarons) of a cation-radical nature with paramagnetic properties (more details below) [36,49]. The subsequent transition of the polymer from the Ox to the Red state results in a significant decrease in the number of polarons, which is evidenced by a significant decrease in the intensity of these satellite bands almost to the level of the complex spectrum.

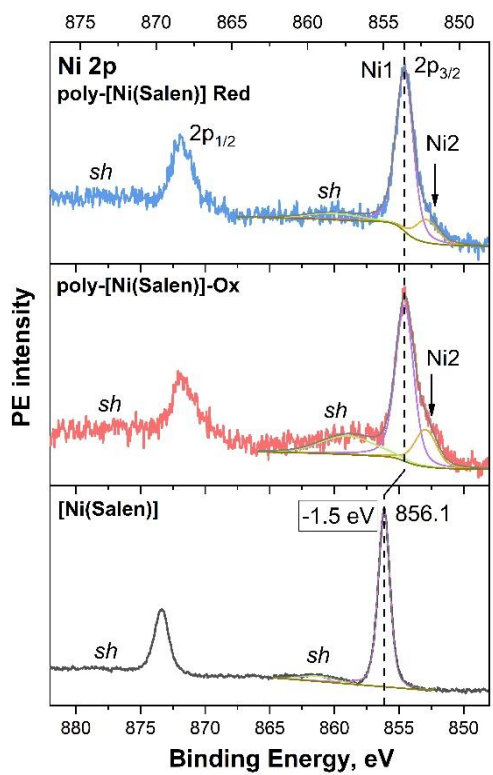

**Figure 4.** Ni 2*p* PE spectra of [Ni(Salen)] and its polymer in different charge (oxidized (Ox) and reduced (Red)) states (*hv* = 1486.6 eV)

Now let us consider the N 1*s* and O 1*s* PE spectra of the complex and its polymer in different charge states, which are shown in Figures 5a and 5b. It is evident that in both systems the N 1*s* spectra contain a single symmetric PE line (Figures 5a). However, in the case of the polymer, this line is somewhat broadened and shifted by -1.7 eV relative to the PE line of the complex. At the same time, the value of this shift is the same for both charge states of the polymer.

When comparing the O 1*s* PE spectra of the [Ni(Salen)] complex and its polymer, a significant broadening of the PE line was found, which clearly indicates the presence of O in several chemically nonequivalent states in the polymer (Figure 5b, Table 3). The performed analysis of the O 1*s* PE spectra suggests that the oxygen atoms in polymers are in two chemical states with binding energies of 530.6 and 532.0 eV, respectively. It is obvious that the most intense O1 component corresponds to oxygen in the [$NiO_2N_2$] coordination centers of the monomer fragments of the polymer. At the same time, it is shifted relative to the O 1*s* spectrum of the complex by -1.5 eV, which can be explained by a change in the valence electron density on the oxygen atoms of the monomers during the formation of the polymer. The origin of the high-energy O2 component can be explained from the standpoint of the formation of polarons particles of radical-cation nature consisting of short d-d dimer fragments cross-linked together through C–C bonds during the electrochemical oxidation of the complex [35]. In this case, according to [23,50], in a number of monomer fragments of such particles, one bond between the nickel atom and oxygen weakens, which is accompanied by a shift in the electron density to the carbon atoms of the salen ligand (see Figure 3a and 5b). As a result, the effective negative charge on the oxygen atoms decreases, which is observed as a shift in the PE line to the region of high binding energies. Interestingly, in the O 1*s* spectrum of the poly-[Ni(Salen)]-Ox, the O2 component has a higher intensity than in the reduced state. In other words, the number of polarons increases with oxidation and decreases with reduction of the polymer, which is also indirectly confirmed by the change in the intensity of the *sh* bands in the Ni 2*p* PE spectra of these samples (Figure 4).

To sum up, it is worth noting that in all the CL PE spectra presented above, a noticeable broadening of the main lines and a low-energy shift of 1.5-1.8 eV are observed upon transition from the complex to the polymer. It is logical to assume that these results reflect the redistribution of the valence electron density between the ligand and nickel atoms in the monomer, caused by a change in their chemical bonds during polymerization. In this case, the greatest shift to the region of low $E_{bin}$ is observed for carbon atoms, which is probably due to the formation of an extended π-conjugated bonding system among the carbon atoms in the polymer.

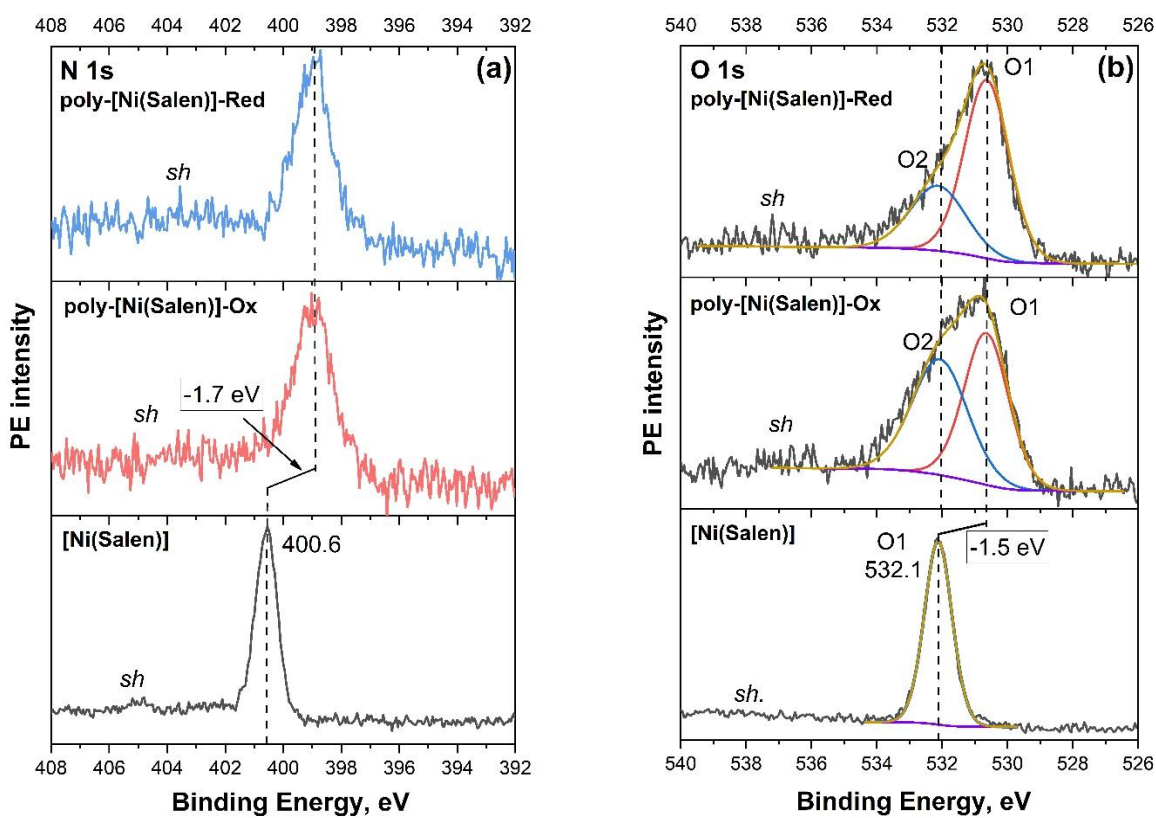

**Figure 5.** (a) N 1*s* and (b) O 1*s* PE spectra of [Ni(Salen)] and its polymer in different charge (oxidized (Ox) and reduced (Red)) states ($hv$ = 1486.6 eV)

**Table 3.** Energy positions of the 1*s* lines of the C, N, O atoms and the 2$p_{3/2}$ line of the Ni atom in the photoelectron spectra of the [Ni(Salen)] complex and its polymer in different charge states. The energy shift of the corresponding band in the polymer relative to the same band in the complex is given in brackets

| Sample | $E_{bin}$, eV | | | | | | |
|---|---|---|---|---|---|---|---|
| | C 1*s* | | O 1*s* | | N 1*s* | Ni 2$p_{3/2}$ | |
| | Cα | Cβ | O1 | O2 | | Ni1 | Ni2 |
| [Ni(Salen)] | 285.8 | 287.3 | 532.1 | - | 400.6 | 856.1 | - |
| poly-[Ni(Salen)]-Ox | 284.0 (-1.8) | 285.5 (-1.8) | 530.6 (-1.5) | 532.1 | 398.9 (-1.7) | 854.6 (-1.5) | 853.0 |
| poly-[Ni(Salen)]-Red | 284.0 (-1.8) | 285.5 (-1.8) | 530.6 (-1.5) | 532.1 | 398.9 (-1.7) | 854.6 (-1.5) | 853.0 |

Finally, let us move on to the consideration of the F 1*s* and B 1*s* PE spectra of the reference compound LiBF$_4$ and the polymer in the oxidized and reduced states (Figure 6). It can be seen that the F 1*s* and B 1*s* PE spectra of the reference compound are characterized by single almost symmetrical lines at the $E_{bin}$ of 687.4 and 196.0 eV, respectively, which are consistent with the energy positions of F and B 1*s* lines in the PE spectra of the powdered LiBF$_4$ [51,52]. However, when moving to the polymer spectra, the position of these lines shifts towards lower binding energies of 685.2 and 192.1 eV, respectively. At the same time, it can be seen that these lines are symmetrical in both polymer cases. All this points to the presence of fluorine and boron in only one chemical (charge) state of BF$_4^-$. The main changes in the spectra of the polymer in different charge states are observed only in the intensity of the F 1*s* and B 1*s* lines: for the polymer in the oxidized state, the intensity of these PE lines increases, while for the reduced state, on the contrary, it decreases. In other words, all this indicates that the concentration of BF$_4^-$ counterions increases during the oxidation of the polymer and decreases during its reduction, which correlates with the data of chemical analysis of the samples studied (Table 1). However, some of the BF$_4^-$ anions probably do not leave the polymer film and are retained in the pores or inside its globules.

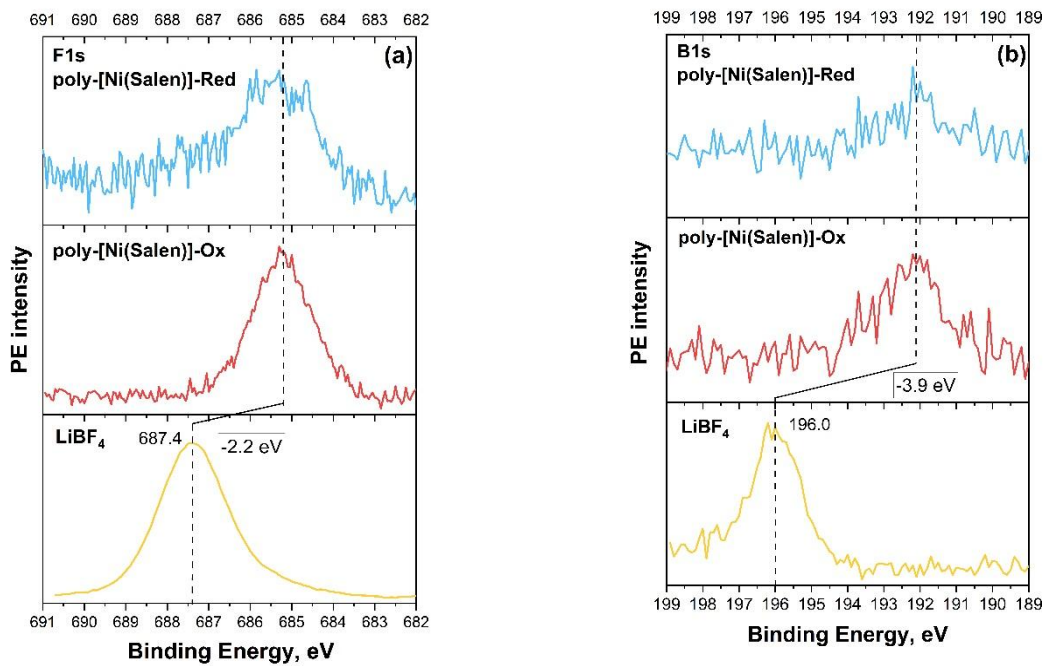

**Figure 6.** (a) F1s and (b) B1*s* PE spectra of [Ni(Salen)] and its polymer in different charge (oxidized (Ox) and reduced (Red)) states (*hv* = 1486.6 eV)

## 3.2 UV PES

To obtain information about the electronic structure of the poly-[Ni(Salen)] valence band, a comparative analysis of the UV PE spectra of the [Ni(Salen)] complex and the poly-[Ni(Salen)] polymer in different charge states was carried out. This analysis was supplemented by the DFT calculations of the electronic structure of the valence band of these systems (Figures 7, 8).

It is clearly seen in Figure 7 that the shapes of the PE spectra of the complex and polymer are characterized by different spectral distributions of PE intensity within the $E_{bin}$ range of 1.8–16.0 eV. Let us first consider the region of $E_{bin}$ from 6.0 to 16.0 eV, where the *c–f* PE bands are located. A detailed comparison of the PE spectra shows that the *c*, *e*, and *f* bands in the polymer spectra are significantly more intense than in the complex spectrum. A previous study of the valence band PE spectrum of the [Ni(Salen)] complex [25,26] showed that the *c–f* bands are primarily associated with occupied σ-MOs, to which the C 2*s*, C 2*p*, O 2*p*, and N 2*p* AOs contribute. Moreover, the C 2*p*σ AOs, due to the greater number of carbon atoms compared to nitrogen and oxygen atoms (one monomer fragment [Ni(Salen)] ($C_{16}H_{14}N_2O_2Ni$) contains 16 carbon atoms versus two nitrogen atoms and two oxygen atoms), make more significant contributions to these bands. Based on the above, it can be assumed that the more intense *c*, *e*, and *f* PE bands in the UV PE spectra of the polymer compared to the complex suggest an increased contribution from of the 2*p*σ AOs of carbon atoms. This is likely a result of new C–C bonds formed during the cross-linking of dimeric fragments in the polymerization process, which is confirmed by the analysis of the C 1*s* PE spectra of the polymer (see Figure 3a). In addition, the UV PE spectra of poly-[Ni(Salen)] reveal a noticeable enhancement of the PE signal in the $E_{bin}$ regions between the *c* and *d* bands, as well as between the *b* and *c* bands, relative to the spectrum of [Ni(Salen)]. This enhancement is more pronounced in the spectrum of the oxidized form (poly-[Ni(Salen)]-Ox) than in its reduced counterpart (poly-[Ni(Salen)]-Red). In other words, it depends on the degree of oxidation of the polymer. Taking into account the earlier results from Section 3.1, it can be assumed that the rise in PE signal intensity within the binding energy range of 6–9 eV is associated with the superposition of electronic states from $BF_4^-$ counterions (adsorbed from the electrolyte during electrochemical polymerization of [Ni(Salen)]) onto the UV PE spectrum of the polymer.

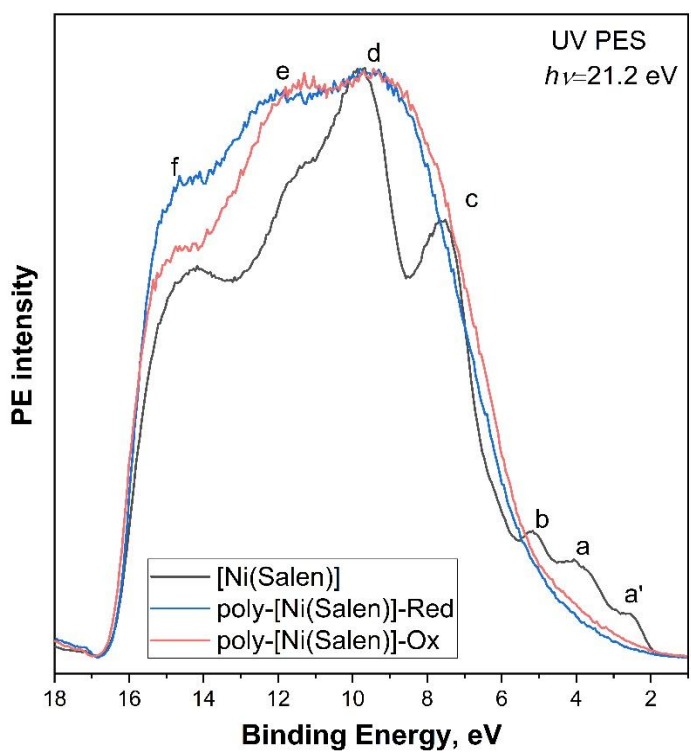

**Figure 7.** UV PE spectra of the [Ni(Salen)] complex (black curve) and the poly-[Ni(Salen)] polymer in the oxidized (red curve) and reduced (blue curve) states, measured at $h\nu$ = 21.2 eV. Lowercase letters *a′, a, b, c, d, e* and *f* denote the PE signals of individual subbands of the valence band. The spectra are normalized to the intensity of the signal maximum in the region of the PE band *d* and plotted on the $E_{bin}$ scale relative to the Fermi level ($E_F$ = 0 eV)

Other differences between the UV PE spectra of the studied systems are observed in the binding energy range of 1.8–6.0 eV. In this region of the [Ni(Salen)] spectrum, there are clear PE bands *a′*, *a* and *b* at 2.5, 3.8 and 5.2 eV, whereas in the polymer spectra this region is characterized by a practically structureless energy distribution of the photoemission signal. This is probably due to the redistribution of the valence electron density between the ligand and nickel atoms in the monomer, caused by a change in their chemical bonding during polymerization (cross-linking of individual dimer fragments and formation of an extended polymer structure). This assumption is consistent with the previously discovered noticeable broadening of the main C 1*s*, N 1*s*, O 1*s*, Ni 2$p_{3/2}$ PE lines and their low-energy shift by 1.5–1.8 eV upon transition from the complex to the polymer (detailed information is presented in Section 3.1, see Figures 3–5).

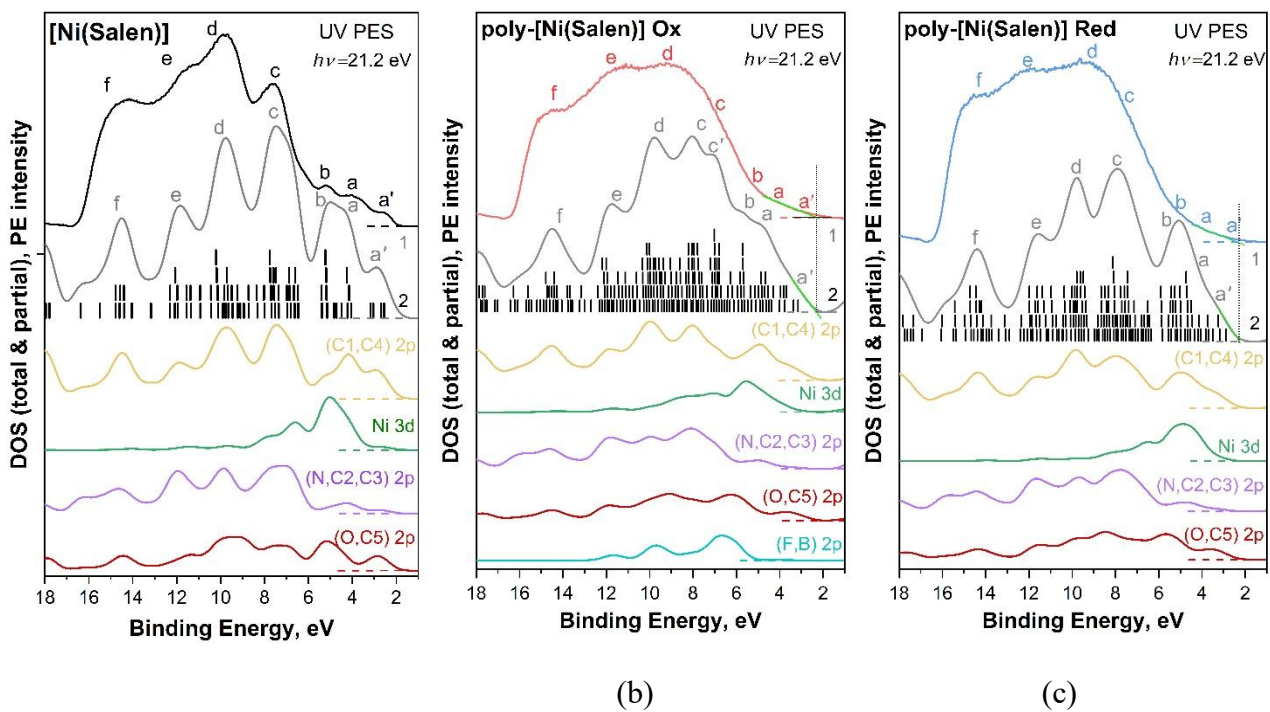

(b) (c)

**Figure 8.** Comparison of the UV PE spectra of the [Ni(Salen)] complex (black curve) (a) and its polymer poly-[Ni(Salen)] in the oxidized (red curve) (b) and reduced states (blue curve) (c) with the energy distributions of the total (1) and partial DOS (2) in the valence band, calculated using DFT. The calculated energy positions of the valence MOs, which are indicated by vertical black lines, were convoluted using a Gaussian contour with FWHM=1.0 eV to make a correct comparison with the experimental spectra

The experimental UV PE spectra and the calculated total and partial DOS spectra of the studied systems for the $E_{bin}$ range from 1.0 to 18 eV are shown in Figure 8. The theoretical curves for [Ni(Salen)] and its polymer are energetically aligned with the PE band *a'* observed in the experimental UV PE spectrum of each of these compounds. At the same time, it can be seen (Figures 8b and 8c) that a direct comparison of the UV PE spectra of the polymers in both charge states and the corresponding total DOS spectra is difficult, since the experimental spectra are characterized by an almost structureless energy distribution of the PE signal in the $E_{bin}$ range of 1.8–6.0 eV. For this reason, the theoretical curves were aligned with the onset of the valence band in the experimental UV PE spectra of poly-[Ni(Salen)]. This onset was defined as the energy value at the point where the extrapolation line (green color) of the signal rise in the 4.5–2 eV region (*b*–*a′*) intersects the background line of the spectrum.

When comparing the total DOS spectra for poly-[Ni(Salen)] and [Ni(Salen)], the first thing that attracts attention is the difference in their spectra in the region of $E_{bin}$ below 4 eV, namely: the *a′* band is absent in the spectra of the polymer, whereas in the spectrum of the [Ni(Salen)] complex, it is clearly distinguishable and has a local maximum (Figure 8, grey curves). In the energy range above the PE band *a′* the shape of the poly-[Ni(Salen)]-Ox spectrum in comparison with that of the complex demonstrates a number of visible differences: the intensity of band *b* decreases compared to band *a* and they resolve energetically to form distinct local maxima, while band *c* is split into two bands *c* and *c′* when passing from the spectrum of the complex to that of poly-[Ni(Salen)]-Ox. At the same time, during the subsequent transition from the spectrum of poly-[Ni(Salen)]-Ox to that of poly-[Ni(Salen)]-Red, the spectral shape of the latter practically returns to the shape of the monomer complex spectrum, with the exception of a slight decrease in the intensity of the band *c* and the absence of the band *a′*.

A detailed examination of the partial DOS spectra for the studied systems allows us to assume that the PE band *a′* in the UV PE spectra of the complex and the polymer in both charge states has the same nature and is due to the photoionization of the HOMO, formed predominantly by the atomic $2p\pi$ orbitals of the carbon atoms (C1 and C4) of the phenolic fragments. The following bands *a* ($E_{bin}$ = 3.8 eV) and *b* ($E_{bin}$ = 5.2 eV), which are well resolved in the PE spectrum of the [Ni(Salen)] complex and are not explicitly observed in the polymer spectra, are largely due to valence MOs with dominant contributions of the Ni $3d$ AOs and the O and N $2p$ AOs of the ligand atoms. The high-energy PE bands *c*, *d*, *e* and *f* in the spectrum of [Ni(Salen)] reflect mainly the MOs formed by $2p\sigma$ AOs of the salen ligand atoms with a small contribution from Ni $3d$ AOs (the largest contribution from these AOs is observed for the *b* band). In turn, in the spectrum of the poly-[Ni(Salen)]-Ox polymer in the energy region responsible for the PE bands *c*, *d* and *e*, an additional contribution from the $2p$-orbitals of the $BF_4^-$ counterion atoms is also observed. This result is a consequence of the superposition of the electronic states of the valence band of the $BF_4^-$ counterions on the electronic states of the polymer spectrum in the oxidized state, which leads to the blurring of the fine structure of the UV PE spectrum of poly-[Ni(Salen)]-Ox. It is worth noting that in the UV PE spectrum of the poly-[Ni(Salen)]-Red in the same energy region an increase in the PE signal is also observed, although less significant. This indicates that during the transition from the oxidized to the reduced state, a certain amount of counterions is retained in the structure of the polymer film, which is consistent with the analysis of the F $1s$ and B $1s$ PE spectra as well as the results of chemical analysis (Figure 6, Table 1).

### 3.3 NEXAFS

Let us begin our consideration with the C 1$s$ NEXAFS spectra of the [Ni(Salen)] complex and its polymer poly-[Ni(Salen)] in the oxidized and reduced states, shown in Figure 9. As demonstrated earlier [25,26], the carbon atoms in the complex are part of two polyatomic fragments (quasimolecules) of the ligand – the phenolic ($C_6H_5OH$) and ethylenediamine ($C_2H_4(NH_2)_2$) groups. Therefore, the absorption bands (resonances) in the [Ni(Salen)] spectrum caused by dipole-allowed transitions of C 1$s$ electrons of these polyatomic groups to unoccupied electron states of the latter are marked differently in Figure 9 with capital letters with asterisks and dashes, respectively. Based on a detailed analysis of the C 1$s$ absorption spectra of the [Ni(Salen)] complex and the $H_2$(Salen) molecule [26], it was found that the low-energy $A'$–$D^*$ and high-energy $E'$–$F^*$ resonances are caused by dipole-allowed transitions of 1$s$ electrons into unoccupied antibonding π- and σ-MOs of the ethylenediamine and phenolic fragments of the salen ligand.

It is clearly seen that, in general the compared spectra are very similar both in terms of the overall spectral distribution of the absorption intensity and in the number and energy positions of the absorption bands. However, there are some significant differences for the absorption resonances in the low-energy region of the polymer spectra compared to that of the complex. The most variations in the polymer spectra are the noticeably lower contrast and broadening of the $B'$, $A^*$, $B^*$ resonances and the almost complete disappearance of the $C^*$ and $D^*$ resonances associated with transitions of C 1$s$ electrons to unoccupied antibonding π-MOs [26]. It should also be noted that in the polymer spectra the ratio of the intensities of the two dominant π-resonances $B'$ and $A^*$ is reversed compared to that of [Ni(Salen)]. As for the high-energy $E'$–$F^*$ σ-resonances, it can be stated that they are quite similar in the spectra of the complex and its polymer in both charge states. This means that the atomic framework of the monomer ligand, provided by the σ-bonding of carbon atoms, changes only slightly during polymerization. Thus, it is obvious that these findings in the C 1$s$ NEXAFS spectra of polymers primarily reflect changes in the local π-electronic structure in the region of the carbon atoms as a result of polymerization of the monomer molecules.

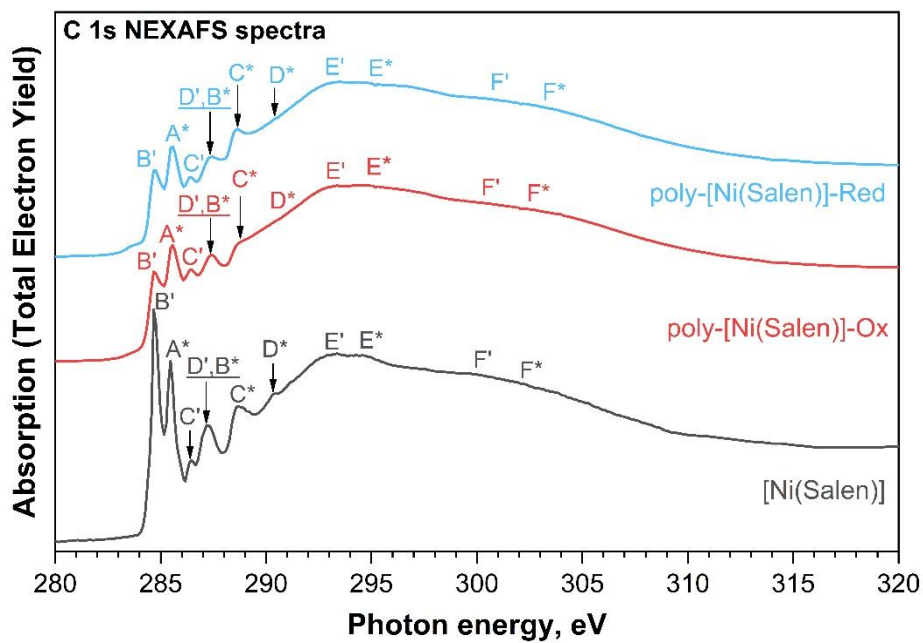

**Figure 9.** C 1$s$ NEXAFS spectra of the [Ni(Salen)] complex and the poly-[Ni(Salen)] polymer (in the oxidized and reduced states) in the photon energy range of 280–320 eV. Capital letters with dashes and asterisks denote absorption bands due to transitions of C 1$s$ electrons of ethylenediamine and phenolic groups to unoccupied electronic states, respectively. The spectra are normalized to the intensity of the continuous absorption jump at a photon energy of 320 eV

The spectral differences in the intensity of the phenolic resonances $A^*$, $B^*$, $C^*$ and $D^*$ upon transition from the spectrum of the [Ni(Salen)] to the spectra of the poly-[Ni(Salen)] indicate a decrease in the contributions of the valence C 2$p$ AOs of phenolic carbon atoms C1 and C5 to the unoccupied antibonding π-MOs (Figures 3b and 9). This can be interpreted as a result of π-bonding (conjugation) of the phenolic fragments of neighboring monomers during polymerization. It is clear that these changes in C 2$p$ contributions to MOs are accompanied by a transfer of electron density between the monomer atoms and, as a result, changes in their charge states. As noted, when examining the XPS and UV PE spectra, this leads to energy shifts of the CL PE lines and their broadening, as well as broadening and changing the intensity of the PE bands of the valence subbands (see Section 3.1).

The main change in the ethylenediamine fragment is a significant decrease in the intensity of the $B'$ band, as well as some weakening of the $C'$ and $D'$ bands, which are associated with the transitions of C 1$s$ electrons to the antibonding π-MO of the –N=C– functional group of the imine. Taking into account the fact that the nitrogen atoms of the ethylenediamine fragment participate in the formation of the [NiO$_2$N$_2$] coordination center, it is logical to associate the discovered features for the $B'$, $C'$, and $D'$ bands with changes in the structure of the coordination center during polymerization of the [Ni(Salen)] monomer fragments.

Finally, it should be noted that the C 1$s$ absorption spectra of both the oxidized and reduced polymers differ little from each other, with only a slight increase in the contrast of the π-resonances $B'$–$C^*$ observed upon going from poly-[Ni(Salen)]-Ox to poly-[Ni(Salen)]-Red.

The N 1$s$ NEXAFS spectra of the [Ni(Salen)] complex and its polymer in the oxidized and reduced states are compared in Figure 10. It is clearly seen that the spectra compared coincide in the number of the main absorption bands $B$–$E$ and their energy positions and also have similar general spectral behavior of the N 1$s$ photoabsorption cross section. When moving from the spectrum of the [Ni(Salen)] complex to that of the polymer in the oxidized state, a noticeable (about 20%) decrease in the relative intensity of the dominant resonance $B$ is observed in the π-region (resonances $B$, $B_1$, $B_2$ [25,26]). This resonance is also slightly broadened and shifted towards higher-energy photons by about 0.15 eV.

In the σ region (resonances $C$, $D$, and $E$ [25,26]) upon transition from the spectrum of the complex to that of the polymer poly-[Ni(Salen)]-Ox, there is a slight weakening of the $D$ resonance intensity. This absorption band was previously assigned in the complex spectrum to the N 1$s$ electron transitions to σ-MOs oriented towards the C2 carbon atoms [26]. Therefore, this change in intensity appears to be related to an alteration in the orientation of the ethylene group within the ethylenediamine fragment, due to a distortion in the geometric structure of the [NiO$_2$N$_2$] coordination center for a number of monomeric molecules during their electrochemical polymerization [35]. It should be noted that this conclusion is consistent with a decrease in the intensity of the π-resonance $B$ in the spectrum of the oxidized polymer.

Figure 10 clearly shows that after the transition from the Ox state to the Red state, the N 1$s$ spectrum of poly-[Ni(Salen)]-Red practically coincides with that of the [Ni(Salen)] complex. The intensity of band $B$ in the spectrum of poly-[Ni(Salen)]-Red increases to its initial value in that of the complex. This spectral similarity suggests the restoration of the ethylene group's original orientation and, consequently, a decrease in distortions within the [NiO$_2$N$_2$] coordination center when the oxidized polymer returns to its neutral state, which is in good agreement with the conclusions of previous work [35].

Overall, the presence of only relatively small differences in the N 1$s$ spectra of the polymer compared to the spectrum of the complex indicates that the imine (–C=N–) functional groups and ethylene (C$_2$H$_4$) fragments in the monomer molecules remain unchanged during the polymerization process.

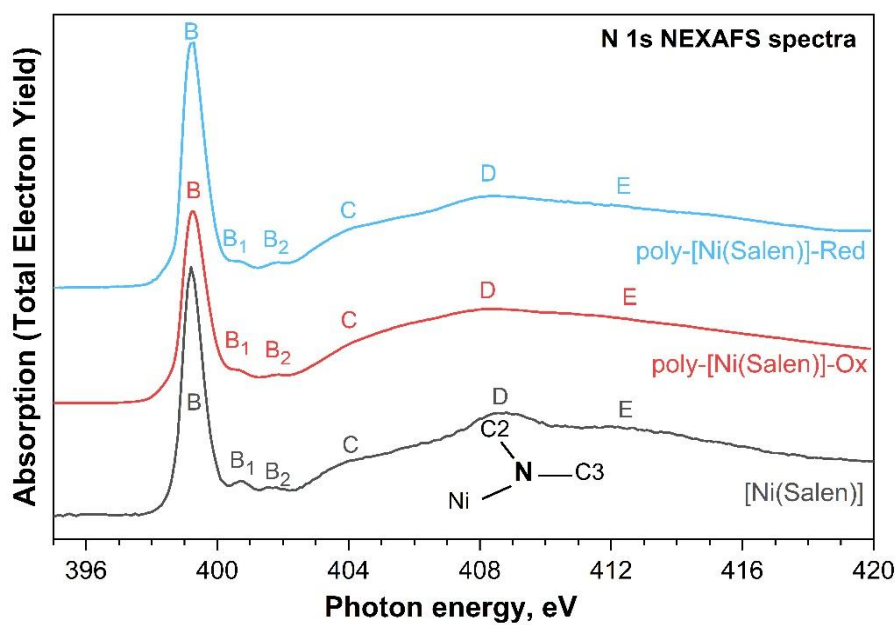

**Figure 10.** N 1$s$ NEXAFS spectra of the [Ni(Salen)] complex and its polymer poly-[Ni(Salen)] (in the oxidized and reduced states) in the photon energy range of 395–420 eV. The spectra are normalized to the intensity of the continuous absorption jump at a photon energy of 420 eV. The inset shows the nearest surroundings of the nitrogen atoms in the monomer fragment (according to Figure 3b)

Let us now consider the O 1$s$ NEXAFS spectra of the [Ni(Salen)] complex and its polymer poly-[Ni(Salen)] in the oxidized and reduced states, shown in Figure 11. First, it should be noted that the spectra being compared exhibit a similar behavior of absorption intensity, with the same fine structure consisting of two absorption resonances $B$ and $B_1$ in the lower energy π-region of the spectrum, and absorption bands $C$, $D_1$, and $E$ at higher photon energies(σ-region). Along with this, it is clearly seen that the fine structure of the O 1$s$ absorption spectrum undergoes noticeable changes upon transition from the complex spectrum to that of oxidized polymer. First of all, this concerns the low-energy region, where a significant decrease in the intensity and broadening of the low-energy π-resonances $B$ and $B_1$ are observed in polymer spectra. It is important to note that the spectra of the polymer in the Ox and Red states differ significantly from each other, namely: the $B$ resonance in the spectrum of the oxidized polymer has a noticeably lower intensity than in the case of the spectrum of the reduced polymer. As shown in [25,26,35], this resonance is associated with the transitions of O, N 1$s$ electrons to the antibonding $\pi e_g$-MO of the [NiO$_2$N$_2$] coordination center, which is formed as a result of covalent π-bonding of the N 2$p\pi$, O 2$p\pi$ and Ni 3$d_{xz,yz}\pi$ AOs and reflects the presence of an additional π-bond between the complexing Ni atom and the ligand atoms N and O in the coordination center. Taking this into account, it can be concluded that the decrease in the intensity of $B$ resonance in the spectrum of the oxidized polymer is likely due to a reduction in the contribution of O 2p orbitals to the antibonding $\pi e_g$-MO. This might indicate a weakening of the π-bond between the oxygen atom and the coordinating nickel atom, which correlates with the observed appearance of the O2 component in the O1s PE spectra of the polymers (Figure 4b). During the transition of the polymer from the oxidized to the neutral state, this π-bonding is partially restored, but not completely. This is manifested by an increase in the intensity of the *B* resonance, which still remains less than in the spectrum of the complex.

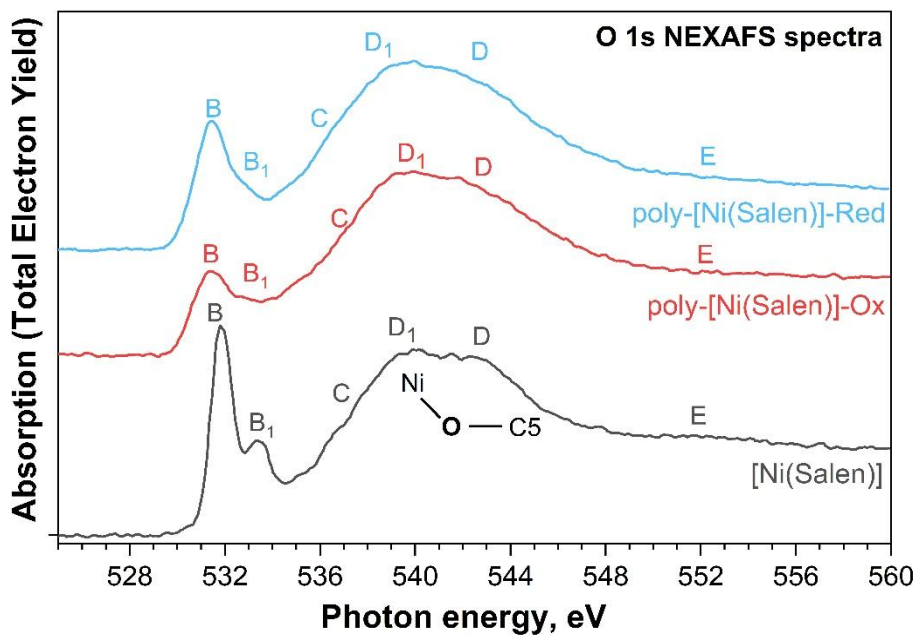

**Figure 11.** O 1*s* NEXAFS spectra of the [Ni(Salen)] complex and its polymer poly-[Ni(Salen)] (in the oxidized and reduced states) in the photon energy range of 525–560 eV. The spectra are normalized to the intensity of the continuous absorption jump at a photon energy of 560 eV. The inset shows the nearest surroundings of the oxygen atoms in the monomer fragment (according to Figure 3b)

The $B_1$ band, which characterizes π-conjugation between oxygen atoms and carbon atoms of the phenyl group [25,26], is well-resolved and clearly visible in the spectrum of the [Ni(Salen)] complex. In the spectra of both poly-[Ni(Salen)], it manifests as a high-energy shoulder of the dominant π-band *B*. In addition, the overall shape of the O 1*s* spectra of the polymers in the π-region is closely similar to the shape of the free $H_2$(Salen) spectrum [26]. This similarity suggests that the molecular field in the region of the C5–O–Ni fragment differs from the planar one in the case of a number of monomers for which the π-bonding between the oxygen and nickel atoms is weakened. As a consequence, the absorption transitions of O 1*s* electrons to the vacant low-energy MOs with O 2p contributions have slightly different energies. This explains the large width of the absorption band B and the absence of a clearly observed band $B_1$ in the spectra of polymers compared to the spectrum of the complex.

In the σ-region of the O 1*s* absorption spectra of the polymers (bands *C*, $D_1$, and *D*), small changes are observed compared to the spectrum of the complex. These changes are mainly due to a decrease in the intensity of the σ-resonance *D*. This effect is likely related to a change in the position of oxygen atoms relative to the plane of the monomer fragment, and determines, in particular, the structural distortions of the coordination center, as noted in previous work [35]. However, in general, the σ-bonding of oxygen atoms in the phenolic groups and the coordination center does not change much upon moving from the complex to the polymer.

Now let us compare the Ni $2p_{3/2}$ spectra of the [Ni(Salen)] complex and its polymer poly-[Ni(Salen)] in the oxidized and reduced charge states (Figure 12a). Despite the overall similarity in the spectra of the monomeric complex and polymers, there is a significant difference between them. It is associated with the appearance of a new low-energy band *A′* in the Ni $2p_{3/2}$ spectra of the polymers, which is most intense in the spectrum of the oxidized polymer. Previously in [35] we showed that this band is most likely associated with transitions of Ni $2p_{3/2}$ electrons to $\sigma b_{1g}$ MOs localized on Ni atoms with a reduced effective charge due to interactions with the counterions of the $BF_4^-$ electrolyte. To confirm the relationship between the appearance of the *A′* band and the presence of electrolyte counterions on the surface of the polymer film, additional studies were conducted on poly-[Ni(Salen)] samples for which $LiClO_4$ salt was used during electrochemical polymerization instead of $LiBF_4$. It is clearly seen (Figure 12b) that when using an electrolyte based on lithium perchlorate, a similar intense band *A′* is also observed in the Ni $2p_{3/2}$ spectra of poly-[Ni(Salen)]-Ox and poly-[Ni(Salen)]-Red. This finding allows us to unambiguously identify the appearance of this band as a result of the transfer of electron density from the $BF_4^-$ ($ClO_4^-$) counterions to the complexing $Ni^{2+}$ cations due to interaction of the polymer nickel atoms and the electrolyte counterions. It is worth noting that for both types of the prepared polymers, upon going from the Ni $2p_{3/2}$ spectra of poly-[Ni(Salen)]-Ox (red and dark yellow curves in Figure 12) to the spectra of poly-[Ni(Salen)]-Red (blue and green curves in Figure 12), the *A′* band does not completely disappear. This result indicates that the counterions do not fully return to the electrolyte during the reduction reaction. Apparently, this is due to the presence of a globular structure of the poly-[Ni(Salen)] polymer films, namely: during film growth, counterions can get between tightly located globules or become locked inside them. Finally, it should be noted that it is the electrolyte counterions that affect the charge state of nickel atoms in the polymer and contribute to weakening of the Ni–O bonds at the stage of oxidation during polymer formation. This conclusion is further supported by the analysis of the O 1*s* PE and NEXAFS spectra (Figures 5b and 11). Thus, we believe that the counterions are responsible for the appearance of the *A′* and Ni2 bands in the Ni $2p_{3/2}$ NEXAFS and PE spectra of the polymer in the oxidized state, respectively (Figures 4 and 12).

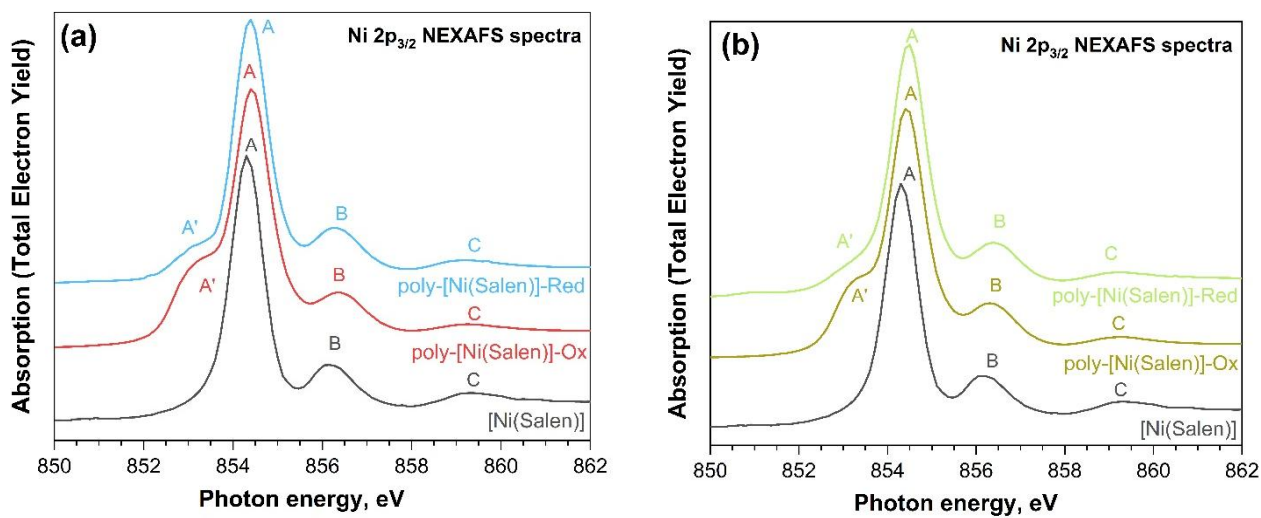

**Figure 12.** Ni $2p_{3/2}$ NEXAFS spectra of the [Ni(Salen)] complex (black curve) and its polymer in the oxidized and reduced charge states prepared using different electrolytes (a) LiBF$_4$/CH$_3$CN (red and blue curves) and (b) LiClO$_4$/CH$_3$CN (dark yellow and green curves). The spectra are normalized to the intensity of the continuous absorption jump at a photon energy of 862 eV

Finally, it is also interesting to compare the energy separation Δ between the Ni1 and Ni2 components in the Ni $2p_{3/2}$ PE spectrum and the *A* and *A'* bands in the Ni $2p_{3/2}$ NEXAFS spectrum of poly-[Ni(Salen)]-Ox. For direct comparison, Figure 13 presents both spectra within the same energy range (850–864 eV). It can be seen that in the Ni $2p_{3/2}$ PE spectrum, the Δ is 0.35 eV higher than in the Ni $2p_{3/2}$ NEXAFS spectrum, which is due to fundamental differences between XPS and NEXAFS. XPS probes the initial electronic state, which is very sensitive to the chemical environment and atomic charge, resulting in pronounced shifts caused by the interaction of nickel atoms with counterions. In contrast, NEXAFS probes unoccupied states in the final state, where the direct influence of counterions is greatly attenuated and screened by the local ligand field of the square-planar [NiO$_2$N$_2$] coordination center. Overall, however, both spectra demonstrate the influence of electrolyte counterions on the charge state of nickel atoms in the oxidized polymer.

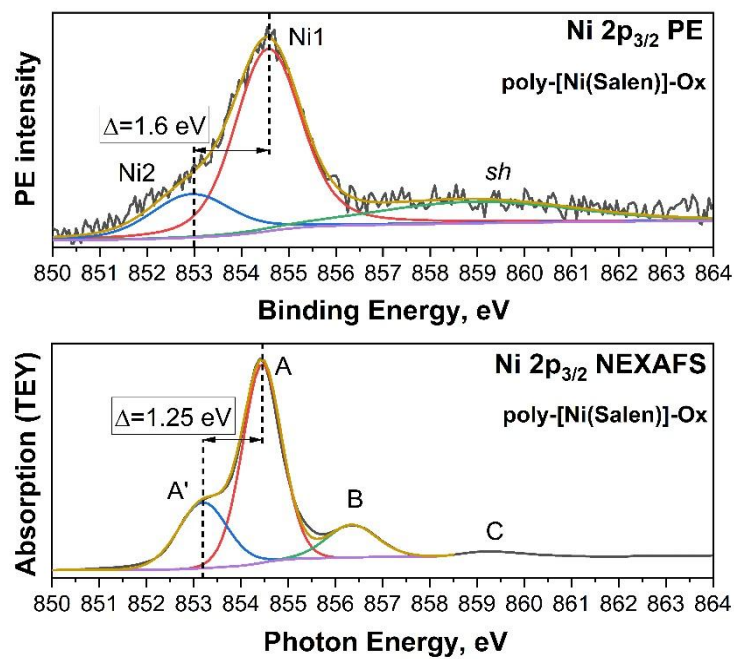

**Figure 13.** Ni $2p_{3/2}$ PE and NEXAFS spectra of the polymer in the oxidized charge state in the energy range of 850–864 eV

Now let us consider the F $1s$ and B $1s$ NEXAFS spectra of the polymer in the oxidized and reduced states, alongside the spectra of the reference compound $LiBF_4$, which are shown in Figure 14. The spectra exhibit a high degree of similarity, both in their general spectral absorption cross-section distribution and in the number of absorption bands. The main difference, however, lies in a high-energy shift of the absorption resonances observed when moving from the $LiBF_4$ to the polymer spectra. Notably, this shift is more pronounced in the B $1s$ spectra than in the F $1s$ spectra of the polymers. This finding suggests that the effective charge on the boron atom undergoes a more significant change than that on the fluorine atoms in the absence of the lithium cation. Taken together, these results confirm the presence of fluorine and boron atoms on the polymer film surface in the form of $BF_4^-$ anions. It is also important to note certain differences in the spectra of poly-[Ni(Salen)]-Ox and poly-[Ni(Salen)]-Red. In particular, the transition from the Ox to the Red state is accompanied by a noticeable decrease in the signal-to-noise ratio, which can be attributed to deteriorated measurement statistics. Despite this, it is important to emphasize that the characteristic absorption bands are still present in the spectrum of poly-[Ni(Salen)]-Red. This observation, combined with the XPS data (Figure 6, Table 1), strongly suggests that some $BF_4^-$ anions remain trapped within the polymer matrix after reduction, most likely located between or within the polymer globules.

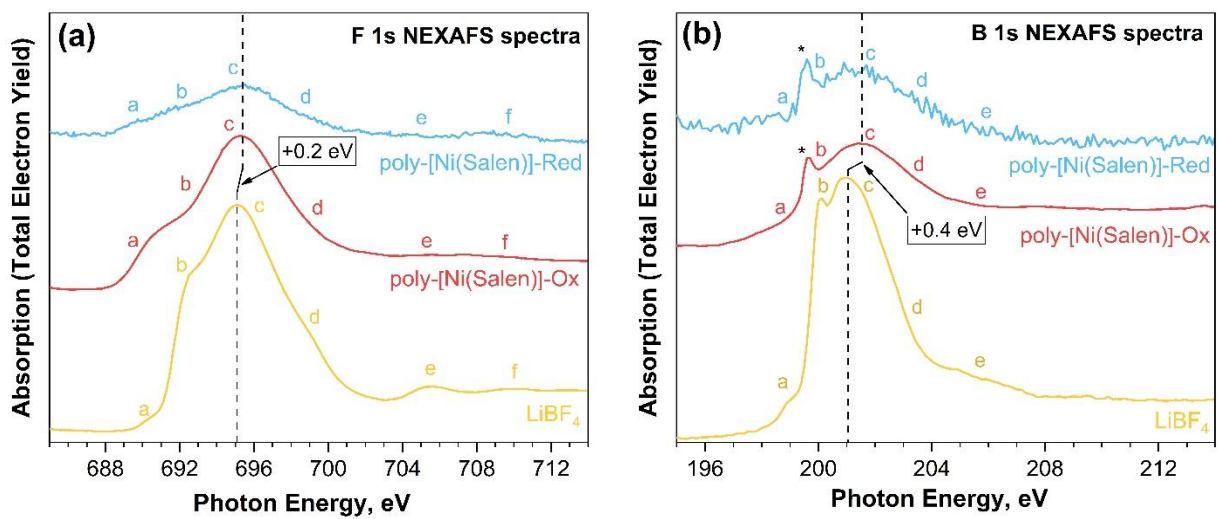

**Figure 14.** (a) F 1*s* and (b) B 1*s* NEXAFS spectra of counterions of the polymer poly-[Ni(Salen)] polymer (in the oxidized and reduced states) and the reference compound LiBF$_4$. In (b), the N 1*s* absorption resonance resulting from the absorption of synchrotron radiation reflected by the diffraction grating in the second order is marked with an asterisk. The F 1*s* and B 1*s* spectra are normalized to the intensity of the continuous absorption jump at a photon energy of 714 and 214 eV, respectively

## 4. Conclusions

A detailed study of the atomic-electronic structure of the conducting polymer poly-[Ni(Salen)] in different (oxidized and reduced) states was conducted using XPS, UV PES, and NEXAFS experimental methods, supplemented by DFT calculations.

The XPS analysis revealed that upon transition from the [Ni(Salen)] complex to the poly-[Ni(Salen)] polymers, significant energy shifts (from -1.5 to -1.8 eV) occur in the 1*s* PE spectra of all ligand atoms (C, N, O) and 2$p_{3/2}$ PE spectra of nickel atoms, accompanied by a noticeable broadening of the spectral lines. This indicates a significant redistribution of valence electron density between the atoms of the monomer fragments in the polymer, caused by changes in their chemical bonds during polymerization. Furthermore, it was shown that the appearance of a new low-energy band in the Ni 2$p_{3/2}$ and a high-energy band in the O 1*s* PE spectra of the oxidized polymer is associated with the weakening of one of the Ni–O bonds. This weakening occurs due to the formation of polarons (charged particles), which consist of short d-d-dimers cross-linked by C–C bonds during the electrochemical oxidation of the complex. The subsequent transition of the polymer from the Ox to the Red state leads to a significant decrease in the number of polarons

The results of DFT calculations demonstrated that the increase in intensity and broadening of the valence subbands in the experimental UV PE spectra upon transition from the complex to the polymer are primarily due to enhancement in the 2*p* contributions of carbon atoms to the occupied MOs. These changes result from the formation of π-conjugation between the phenyl groups of neighboring monomers during polymerization. Additionally, it was shown that the electronic states of the $BF_4^-$ counterions overlap with those of the polymer in the $E_{bin}$ range from 6.0 to 12 eV, which contributes to the blurring of the fine structure of the UV PE spectra of the polymers.

NEXAFS spectroscopy provided additional insight into the electronic structure changes during the transition from the complex to its polymers. A detailed analysis of the N 1$s$ NEXAFS spectra of the studied systems clearly revealed that the ethylenediamine fragment in the monomer molecules is largely retained during polymerization. On the contrary, in the C 1$s$ NEXAFS spectra of polymers, a decrease in the intensity of the main π-absorption resonances associated with phenolic fragments was detected, indicating their direct participation in the polymerization process. In turn, in the low-energy region of the O 1$s$ NEXAFS spectra of the polymers, a decrease in the intensity and contrast of the main absorption resonance was revealed. This is attributed to a reduced contribution of O 2$p$ orbitals to the antibonding $\pi e_g$-MO. Moreover, during the transition of the polymer from the Ox to the Red state, this π-bond is partially restored, but not completely. Finally, it was confirmed that the new band $A'$ observed in the Ni 2$p_{3/2}$ NEXAFS spectrum of the oxidized polymer is clearly related to the interaction of nickel atoms with the counterions of the electrolyte.

Based on XPS and NEXAFS analysis, it was also demonstrated that electrolyte counterions on the polymer film surface are present in the form of $BF_4^-$. These anions significantly affect the charge state of nickel atoms, thereby contributing to the weakening of Ni–O bonds during the electrochemical oxidation of complex fragments in the polymerization process. Furthermore, some of these anions do not leave the polymer film during its reduction from the oxidized state but are instead retained within its pores or globules. As a result, the spectral features characteristic of the oxidized form of the polymer – observed in the O 1$s$ PE and NEXAFS, as well as Ni 2$p$ PE and NEXAFS spectra – do not completely disappear upon transition to the reduced form.

**Acknowledgment**

The authors are grateful to the directorates and administrative staff of the "KISI-Kurchatov" SR source, the Research Park of St. Petersburg State University, as well as the coordinators of the Russian-German laboratory BESSY II.

**Author Contributions**

The manuscript was written through contributions of all authors. All authors have given approval to the final version of the manuscript.

**Funding Sources**

This research was funded by the Russian Science Foundation (grant no. 21-72-10029). The XPS part of the research was funded separately by grant no. 25-21-00376 from the same foundation.